\documentclass[fleqn,twoside]{article}
\usepackage{espcrc2,epsf,latexsym}

\newcommand{\half}{\mbox{\small $\frac{1}{2}$}}          % 1/2
\newcommand{\third}{\mbox{\small $\frac{1}{3}$}}         % 1/3
\newcommand{\twelfth}{\mbox{\small $\frac{1}{12}$}}      % 1/12
\newcommand{\msbar}{\mbox{\tiny $\overline{MS}$}}        % Msbar
\newcommand{\ripmom}{\mbox{\tiny $R\!I^\prime\!\!-\!\!M\!O\!M$}} % RIp-MOM
\newcommand{\rgi}{\mbox{\tiny $R\!G\!I$}}                % RGI
\newcommand{\trbpt}{\mbox{\tiny $T\!R\!B\!\!-\!\!P\!T$}} % TRB-PT
\newcommand{\lat}{\mbox{\tiny $L\!A\!T$}}                % LAT
\newcommand{\con}{\mbox{\tiny $C\!O\!N$}}                % CON
\newcommand{\plaq}{\Box}                                 % PLAQ
\newcommand{\born}{\mbox{\tiny $B\!O\!R\!N$}}            % BORN
\newcommand{\bare}{\mbox{\tiny $B\!A\!R\!E$}}            % bare
\def\lsim{\mathrel{\rlap{\lower4pt\hbox{\hskip1pt$\sim$}}
    \raise1pt\hbox{$<$}}}                % less than or approx. symbol
\def\gsim{\mathrel{\rlap{\lower4pt\hbox{\hskip1pt$\sim$}}
    \raise1pt\hbox{$>$}}}                % greater than or approx. symbol

\title{
       \vspace{-3.65cm}                                     % 
       {\normalsize DESY 06-004}      \\[-0.2cm]            % for preprint
       {\normalsize Edinburgh 2006/02}\\[-0.2cm]            % for preprint
       {\normalsize Liverpool LTH 686}\\[-0.2cm]            % for preprint
       {\normalsize February 2006}  \\                      % for preprint
       \vspace{1.78cm}                                      % for 3 preprint #
       Determining the strange quark mass for 2-flavour QCD%
            \thanks{Talk given by R. Horsley at the
                    Workshop on Computational Hadron Physics,
                    (Nicosia, Cyprus, September '05).}}    % for preprint

\author{M. G\"ockeler%
           \address{Institut f\"ur Theoretische Physik, Universit\"at
                    Regensburg, D-93040 Regensburg, Germany},
        R. Horsley%
           \address{School of Physics,
                    University of Edinburgh, Edinburgh EH9 3JZ, UK},
        A.~C. Irving%
           \address{Department of Mathematical Sciences,
                    University of Liverpool, Liverpool L69 3BX, UK},
        D. Pleiter%
           \address{John von Neumann Institute NIC / DESY Zeuthen,
                    D-15738 Zeuthen, Germany},
        P.~E.~L. Rakow%
           $^{\rm c}$,
        G. Schierholz%
           $^{\rm d,}$%
           \address{Deutsches Elektronen-Synchrotron DESY,
                    D-22603 Hamburg, Germany},
        H. St\"uben%
           \address{Konrad-Zuse-Zentrum f\"ur Informationstechnik Berlin,
                    D-14195 Berlin, Germany}
        and
        J.~M. Zanotti%
           $^{\rm b}$
        \\ -- {\it QCDSF--UKQCD} Collaboration --}

\begin{document}

%----------------------------------------------------------------------------

\begin{abstract}
   Using the $O(a)$ Symanzik improved action an estimate is given for
   the strange quark mass for unquenched ($n_f=2$) QCD.
   The determination is via the axial Ward identity (AWI) and includes
   a non-perturbative evaluation of the renormalisation constant.
   Numerical results have been obtained at several lattice spacings,
   enabling the continuum limit to be taken. Our results indicate a value
   for the strange quark mass (in the $\overline{MS}$-scheme
   at a scale of $2 \, \mbox{GeV}$) in the range
   $100$ -- $130 \, \mbox{MeV}$. A comparison is also made with
   other recent lattice determinations of the strange quark mass
   using dynamical sea quarks.
 \end{abstract}

% typeset front matter (including abstract)
\maketitle

% reset footnote counter
\setcounter{footnote}{0}

%----------------------------------------------------------------------------

\section{INTRODUCTION}

Lattice methods allow, in principle, the complete `ab initio' calculation
of the fundamental parameters of QCD, such as quark masses. 
However quarks are not directly observable, being confined in hadrons
and are thus not asymptotic states. So to determine their mass necessitates
the use of a non-perturbative approach -- such as lattice QCD or
QCD sum rules. In this brief article, we report on our recent results
for the strange quark mass for $2$-flavour QCD in the $\overline{MS}$-scheme
at a scale of $2 \, \mbox{GeV}$, $m_s^{\msbar}(2\, \mbox{GeV})$
(further details can be found in \cite{gockeler05a}).

The present phenomenological status is summarised by the Particle
Data Group in \cite{eidelman04a}
giving an estimate for the strange quark mass of
$80\,\mbox{MeV} < m_s^{\msbar}(2\, \mbox{GeV}) < 130\,\mbox{MeV}$.
This is a large band, and it is be hoped that lattice computations
will reduce this significantly in the coming years.

%----------------------------------------------------------------------------

\section{RENORMALISATION GROUP \\ INVARIANTS}
\label{rgi}

Being confined, the mass of the quark, $m^{\cal S}_q(M)$, needs to be
defined by giving a scheme, ${\cal S}$ and scale $M$,
\begin{equation}
   m_q^{\cal S}(M) = Z_m^{\cal S}(M) m_q^{\bare} \,,
\end{equation}
and thus we need to find both the bare quark mass and
the renormalisation constant. An added complication is that the
$\overline{MS}$-scheme is a perturbative scheme, while more natural
schemes which allow a non-perturbative definition of the renormalisation
constants have to be used. It is thus convenient to first define a
(non-unique) renormalisation group invariant (RGI) object,
which is both scale and scheme independent by
\begin{equation}
   m_q^{\rgi} \equiv \Delta Z_m^{\cal S}(M) m^{\cal S}(M)
               \equiv Z_m^{\rgi} m_q^{\bare} \,,
\label{mrgi_msbar}
\end{equation}
where we have
\begin{eqnarray}
   \lefteqn{[\Delta Z_m^{\cal S}(M)]^{-1} \equiv
     \left[ 2b_0 g^{\cal S}(M)^2 \right]^{- {d_{m0}\over 2b_0}} \times} & &
                                           \nonumber     \\
        & & \exp{\left\{ \int_0^{g^{\cal S}(M)} d\xi
               \left[ {\gamma^{\cal S}_m(\xi)
                          \over \beta^{\cal S}(\xi)} +
                      {d_{m0}\over b_0 \xi} \right] \right\} } \,.
\label{deltam_def}
\end{eqnarray}
The $\beta^{\cal S}$ and $\gamma^{\cal S}_m$ functions (with leading
coefficients $-b_0$, $d_{m0}$ respectively) are known
perturbatively up to a certain order%
\footnote{Analgous definitions also hold for other operators, see
section~\ref{lattice}.}. In the $\overline{MS}$ scheme
the first four coefficients are known, \cite{vanritbergen97a,vermaseren97a},
and this is also true for the $\rm{RI}^\prime$-MOM scheme
\cite{martinelli94a,chetyrkin99a} (which is a suitable scheme for both
perturbative and non-perturbative, NP, applications, see
section~\ref{lattice}). Note that in the $\rm{RI}^\prime$-MOM scheme
we also choose to expand the $\beta^{\cal S}$ and $\gamma^{\cal S}_m$
functions in terms of $g^{\msbar}$, \cite{chetyrkin99a};
other choices, of course, are also possible.

\begin{figure}[t]
   \begin{center}
      \epsfxsize=7.00cm \epsfbox{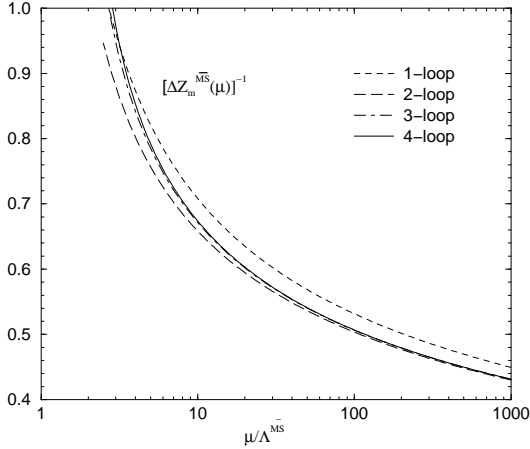}
   \end{center}
   \vspace*{-0.35in}
   \caption{\footnotesize{
            One-, two-, three- and four-loop results for
            $[\Delta Z_m^{\msbar}(\mu)]^{-1}$ in units of
            $\Lambda^{\msbar}$.}}
   \vspace*{-0.15in}
   \label{fig_Del_Zm_MSbar_nf2_muolam}
\end{figure}

\begin{figure}[t]
   \begin{center}
      \epsfxsize=7.00cm \epsfbox{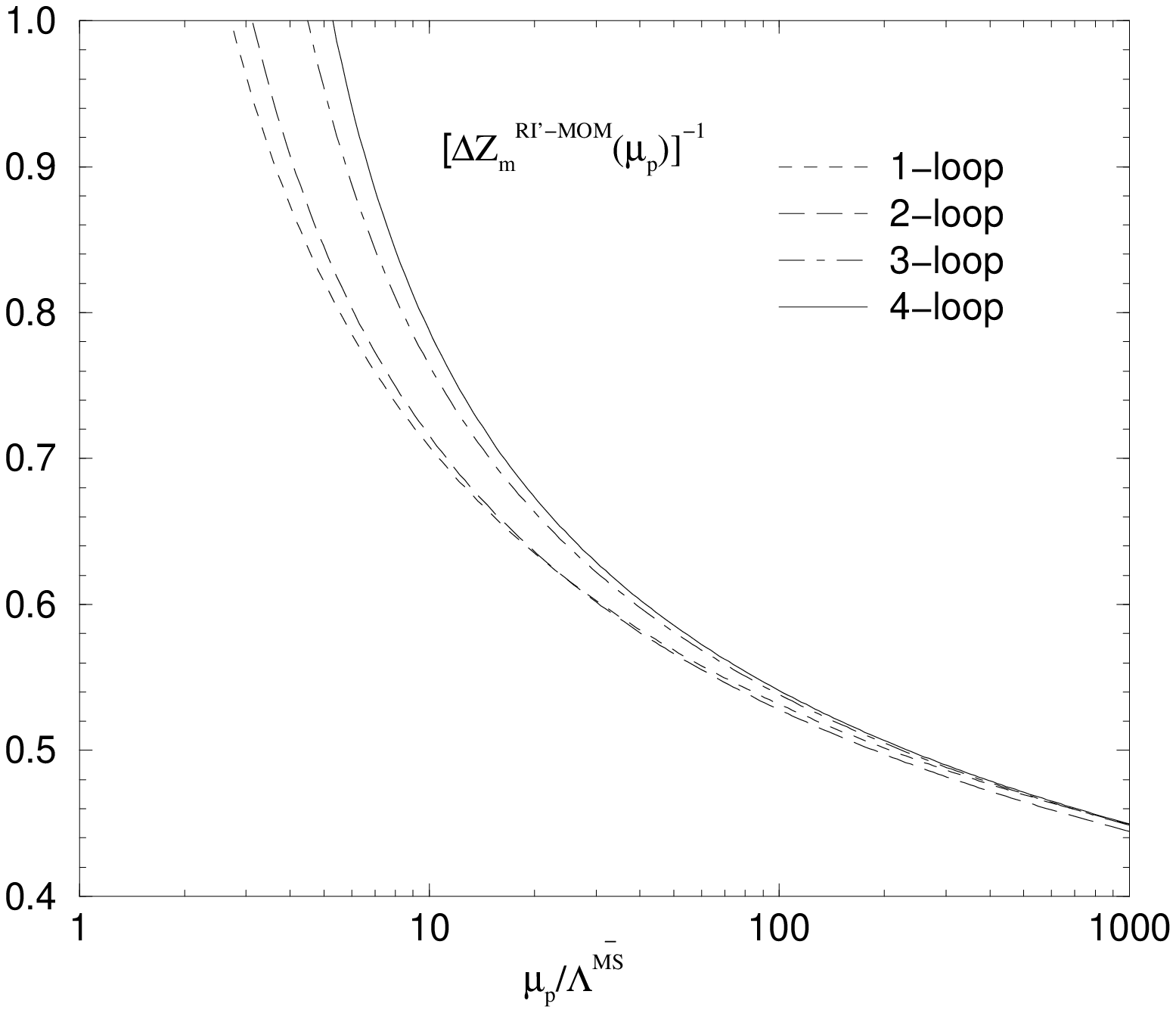}
   \end{center}
   \vspace*{-0.35in}
   \caption{\footnotesize{
            One-, two-, three- and four-loop results for
            $[\Delta Z_m^{\ripmom}(\mu_p)]^{-1}$ in units of
            $\Lambda^{\msbar}$.}}
   \vspace*{-0.30in}
   \label{fig_Del_Zm_MOMms_nf2_muolam}
\end{figure}

In Figs.~\ref{fig_Del_Zm_MSbar_nf2_muolam} and
\ref{fig_Del_Zm_MOMms_nf2_muolam}
we show the results of solving eq.~(\ref{deltam_def}) as a function
of the scale $M \equiv \mu$ and $M \equiv \mu_p$ for both the
$\overline{MS}$ and $\rm{RI}^\prime$-MOM schemes respectively.
We hope to use these (perturbative) results in a region where
perturbation theory has converged. $2 \, \mbox{GeV}$ corresponds to
$\mu/\Lambda^{\msbar} \sim 8$, where it would appear that the expansion
for the $\overline{MS}$-scheme has converged; for the $\rm{RI}^\prime$-MOM
scheme using a higher scale is safer (which is chosen in practice).
However, when the RGI quantity has been determined we can then easily
change from one scheme to another.
Of course these scales are in units of $\Lambda^{\msbar}$ which is
awkward to use: the standard `unit' nowadays is the force scale $r_0$.
To convert to this unit, we use the result for $r_0\Lambda^{\msbar}$
as given in \cite{gockeler05b}. Of course, we must also give the
physical scale. A popular choice is $r_0 = 0.5\,\mbox{fm}$, but
there is some variation in possible values, see section~\ref{results}.

%----------------------------------------------------------------------------

\section{CHIRAL PERTURBATION THEORY}

We have generated results for $n_f = 2$ degenerate sea quarks, together
with a range of valence quark masses. Chiral perturbation theory,
$\chi$PT, has been developed for this case, \cite{bernard93a,sharpe97a}.
We have manipulated the structural form of this equation to give
an ansatz of the form
\begin{eqnarray}
   \lefteqn{r_0m_s^{\rgi} =}
       & &
                                           \nonumber \\
       & & c^{\rgi}_a 
               \left[ (r_0m_{K^+})^2 + (r_0m_{K^0})^2 - (r_0m_{\pi^+})^2 
               \right] +
                                           \nonumber \\
       & & (c^{\rgi}_b-c^{\rgi}_d)
              \left[(r_0m_{K^+})^2 + (r_0m_{K^0})^2 \right] \times
                                           \nonumber \\
       & & \hspace*{0.75in} (r_0m_{\pi^+})^2 +
                                           \nonumber \\
       & & \half (c^{\rgi}_c+c^{\rgi}_d)
                \left[(r_0m_{K^+})^2 + (r_0m_{K^0})^2 \right]^2 -
                                           \nonumber \\
       & & (c^{\rgi}_b+c^{\rgi}_c)(r_0m_{\pi^+})^4 -
                                           \nonumber \\
       & & c^{\rgi}_d \left[(r_0m_{K^+})^2 + (r_0m_{K^0})^2 \right] \times
                                           \nonumber \\
       & &  \hspace*{0.15in}
               \left[ (r_0m_{K^+})^2 + (r_0m_{K^0})^2 - (r_0m_{\pi^+})^2
               \right] \times
                                           \nonumber \\
       & &  \hspace*{0.15in}
                 \ln \left( (r_0m_{K^+})^2 + (r_0m_{K^0})^2 - (r_0m_{\pi^+})^2
                     \right) +
                                           \nonumber \\
       & & c^{\rgi}_d (r_0m_{\pi^+})^4 \ln (r_0m_{\pi^+})^2 \,,
\label{strange_general}
\end{eqnarray}
where
\begin{eqnarray}
   \lefteqn{{ r_0 m_q^{\rgi} \over (r_0 m_{ps})^2 } =  c^{\rgi}_a +}
     & &
                                           \nonumber \\
     & & c^{\rgi}_b (r_0 m^S_{ps})^2
                    + c^{\rgi}_c (r_0m_{ps})^2 +
                                           \nonumber \\
     & & c^{\rgi}_d \left( (r_0 m^S_{ps})^2 - 2(r_0 m_{ps})^2 
                        \right) \ln (r_0 m_{ps})^2 \,.
\label{strange_degenerate}
\end{eqnarray}
$m_{ps}$, $m_{ps}^S$ are the valence and sea pseudoscalar masses 
respectively (both using mass degenerate quarks). The first term is
the leading order, LO, result in $\chi$PT while the
remaining terms come from the next non-leading order, NLO, in $\chi$PT.
Thus we see from eq.~(\ref{strange_degenerate}) that to NLO, we can
first determine $c_a^{\rgi}$ and $c_i^{\rgi}$, $i = b, c, d$
using pseudoscalar mass {\it degenerate} quarks and then simply
substitute them in eq.~(\ref{strange_general}).

%----------------------------------------------------------------------------

\section{THE LATTICE APPROACH}
\label{lattice}

Approaches to determining the quark mass on the lattice are to
use the vector Ward identity, VWI (see e.g.\ \cite{gockeler04a}),
where the bare quark mass is given in terms of the hopping parameter by%
\footnote{This is valid for both valence and sea quarks. $\kappa^S_{qc}$ is
defined for fixed $\beta$ by the vanishing of the pseudoscalar
mass, i.e.\ $m_{ps}(\kappa^S_{qc},\kappa^S_{qc}) = 0$. $\kappa^S_{qc}$
has been determined in \cite{gockeler04a}.}
\begin{equation}
   m_q = {1 \over 2a}
               \left( {1\over \kappa_q} -  {1\over \kappa^S_{qc}} \right) \,,
\end{equation}
or the axial Ward identity, AWI, which is the approach employed here.
Imposing the AWI on the lattice for mass degenerate quarks, we have
\begin{equation}
   \partial_\mu {\cal A}_\mu = 2\widetilde{m}_{q} {\cal P} + O(a^2) \,,
\label{pcac_lattice}
\end{equation}
and ${\cal A}$ and ${\cal P}$ are the $O(a)$ improved%
\footnote{The improvement term to the axial current, $\partial_\mu P$
together with improvement coefficient $c_A$, \cite{dellamorte05a}
has been included. The mass improvement terms, together
with their associated difference in improvement coefficients, $b_A$, $b_P$
appear to be small and have been ignored here.}
unrenormalised axial current and pseudoscalar density respectively
and $\widetilde{m}_{q}$ is the AWI quark mass.
So by forming two-point correlation functions
with ${\cal P}$ in the usual way, this bare quark mass can be determined
\begin{equation}
   a\widetilde{m}_{q}
      \stackrel{t\gg 0}{=}
       { \langle
            \partial^{\lat}_4 {\cal A}_4(t){\cal P}(0)
          \rangle \over
            2 \langle {\cal P}(t){\cal P}(0) \rangle } \,.
\label{correlation_mq}
\end{equation}
We have found results for four $\beta$-values: 5.20, 5.25, 5.29, 5.40,
each with several (three or more) sea quark masses and a variety of
valence quark masses, \cite{gockeler05a}.

Furthermore upon renormalisation we have
\begin{equation}
   {\cal A}^{\cal R}_\mu = Z_A {\cal A}_\mu \,,  \qquad
   {\cal P}^{\cal S}(M)  = Z^{\cal S}_P(M) {\cal P} \,,
\end{equation}
giving from eqs.~(\ref{mrgi_msbar}) and (\ref{pcac_lattice})
\begin{equation}
   Z^{\rgi}_{\tilde{m}} = \Delta Z^{\cal S}_{\tilde{m}}(M)
                              {Z_A \over Z^{\cal S}_P(M)} \,.
\end{equation}

\begin{figure*}[t]
   \hspace{1.00in}
      \epsfxsize=10.00cm
         \epsfbox{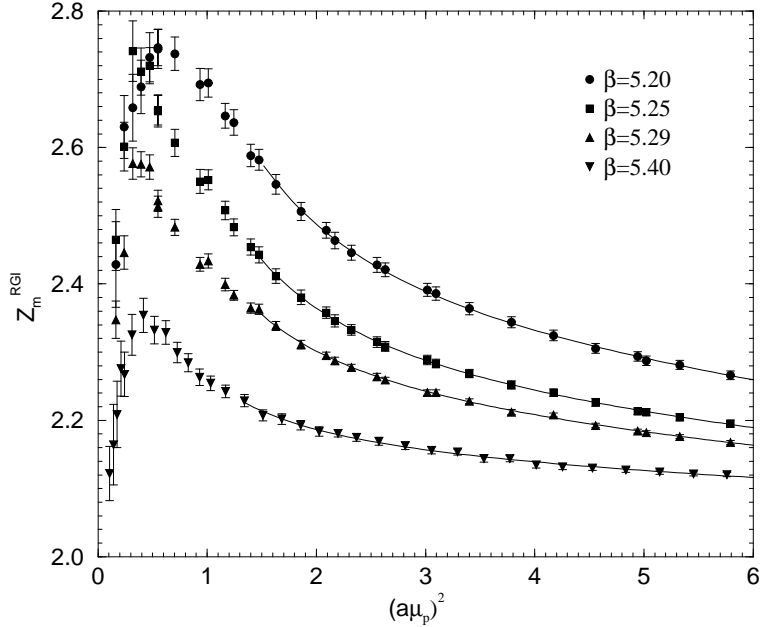}
   \vspace*{-0.35in}
   \caption{\footnotesize{
            $Z_{\tilde{m}}^{\rgi}$ for $\beta = 5.20$ (filled circles),
            $\beta = 5.25$ (filled squares), $\beta = 5.29$ (filled
            upper triangles), $\beta = 5.40$ (filled lower triangles)
            together with fits
            $F(a\mu_p) = r_1 + r_2(a\mu_p)^2 + r_3/(a\mu_p)^2$.}}
%   \vspace*{-0.25in}
   \label{fig_Zm_rgi_b5p20+b5p25+b5p29+b5p40_ap2_050406}
\end{figure*}

As mentioned before, we use the $\rm{RI}^\prime$-MOM scheme,
\cite{martinelli94a}. This scheme considers amputated Green's functions
(practically in the Landau gauge) with an appropriate operator insertion,
here either $A$ or $P$. The renormalisation point is fixed at some
momentum scale $p^2 = \mu_p^2$, and thus we have
\begin{eqnarray}
   \lefteqn{Z_O^{\ripmom}(\mu_p)  =}
       & &
                                           \nonumber \\
       & & \left. { Z_q^{\ripmom}(p) \over
                        \twelfth \mbox{tr}
                           \left[\Gamma_O(p)
                                 \Gamma_{O,\born}^{-1}(p)
                           \right] } \right|_{p^2=\mu_p^2} \,,
\end{eqnarray}
where $\Gamma_O$ are one-particle irreducible (1PI) vertex functions,
and $Z_q$ is the wave-function renormalisation. (Our variation of
the implementation of this method is described in \cite{gockeler98a}.)
This determines $Z_A$ and $Z_P^{\ripmom}$, from which a chiral
extrapolation, here using the sea quarks only, may be made to the
chiral limit. For $Z_A$  we make a linear extrapolation in $am_q$,
$Z_A = A_A + B_A am_q$, while for $Z_P^{\ripmom}$ we must first
subtract out a pole in the quark mass, \cite{cudell98a}, which occurs due
to chiral symmetry breaking. We thus make a fit of the form
$(Z_P^{\ripmom})^{-1} = A_P + B_P / am_q$.
We now have all the components, namely $A_A$, $A_P$ and
$\Delta Z^{\ripmom}_{\tilde{m}}$ necessary to compute $Z_m^{\rgi}$
and hence $r_0m_q^{\rgi}$. In
Fig.~\ref{fig_Zm_rgi_b5p20+b5p25+b5p29+b5p40_ap2_050406}
we show $Z_{\tilde{m}}^{\rgi}$ for $\beta = 5.20$, $5.25$, $5.29$ and
$5.40$ . These should be independent of the scale $a\mu_p$
at least for larger values. This seems to be the case, we make a
phenomenological fit to account for residual effects.

%----------------------------------------------------------------------------

\section{COMPARISON OF $Z_{\tilde{m}}^{\rgi}$ WITH \\ OTHER METHODS}
\label{comparison_ti}

As many computations of the strange quark mass have used
tadpole improved perturbation theory together with a boosted
coupling constant for the determination of the renormalisation constant,
it is of interest to compare our results obtained in the previous
section with this approach. Our variation of this method,
tadpole-improved renormalisation-group-improved boosted perturbation theory
or TRB-PT, is described in \cite{gockeler04b}. Regarding
the lattice as a `scheme', then from eq.~(\ref{mrgi_msbar}) we can write
\begin{equation}
   m_q^{\rgi} = \Delta Z_{\tilde{m}}^{\lat}(a) \tilde{m}_q(a) \,,
\end{equation}
where the renormalisation-group-improved $\Delta Z_{\tilde{m}}^{\lat}(a)$ 
is given  by eq.~(\ref{deltam_def}). Furthermore in this `lattice' scheme,
we choose to use $g_{\plaq}^2 = g_0^2/u_{0c}^4$ where
$u_0^4 = \langle \third \mbox{Tr}U^{\plaq}\rangle$
($U^{\plaq}$ being the product of links around an elementary plaquette)
rather than $g_0$, as series expansions in $g_{\plaq}$ are believed
to have better convergence. This is boosted perturbation theory.
(We shall use chirally extrapolated plaquette values as determined
in \cite{gockeler05a} at our $\beta$-values and so we add a subscript
`$c$' to $u_0$.) In the tadpole-improved method, noting that
renormalisation constants for operators with no derivatives are
$\sim u_{0c}$, which indicates that $Z^{\rgi}_{\tilde{m}}u^{-1}_{0c}$ will
converge faster then $Z^{\rgi}_{\tilde{m}}$ alone we re-write
eq.~(\ref{deltam_def}) in the two loop approximation as%
\footnote{The TRB-PT subscript in brackets is there only to distinguish
the results from those obtained in section~\ref{lattice}.}
\begin{eqnarray}
   Z^{\rgi(\trbpt)}_{\tilde{m}}
      &\equiv& \Delta Z^{\lat}_{\tilde{m}}(a)
                                                 \label{DelZ_lat_TI}  \\
      &=& u_{0c}
            \left[ 2 b_0 g^2_{\plaq} \right]^{d_{m0}\over 2b_0}
            \left[ 1 + {b_1 \over b_0} g^2_{\plaq}
            \right]^{q_1}  \,,
                                                 \nonumber
\end{eqnarray}
where $q_1 = (b_0 d_{\tilde{m}1}^{\lat} - b_1 d_{m0})/(2 b_0 b_1) +
(p_1 / 4 )(b_0 / b_1)$ with $p_1 = \third$ being the first coefficient
in the expansion of $u_{0c}$. $d_{\tilde{m}1}^{\lat}$ may be found by
relating the (known) perturbative result for $Z_{\tilde{m}}^{\msbar}$
to $\Delta Z^{\lat}_{\tilde{m}}$.

\begin{figure}[t]
   \epsfxsize=7.00cm 
      \epsfbox{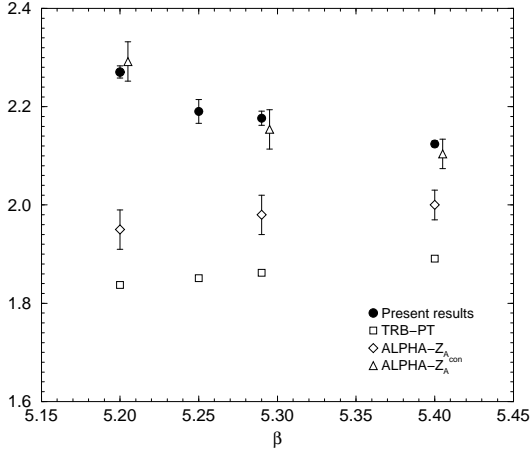}
   \vspace*{-0.25in}
   \caption{\footnotesize{
            $Z^{\rgi}_{\tilde{m}}$ versus $\beta$. The black circles
            are the results from section~\ref{lattice}, while
            the open squares are the TRB-PT results. Furthermore
            the open diamonds and triangles are the NP results from
            \cite{dellamorte05a}, using the two different results
            for the axial renormalisation constant, \cite{dellamorte05c}.
            (The empty triangle results have been slightly displaced
            for clarity.)}}
   \vspace*{-0.20in}
   \label{fig_Zrgi_mtwid_beta_051024}
\end{figure}

In Fig.~\ref{fig_Zrgi_mtwid_beta_051024} we plot $Z^{\rgi}_{\tilde{m}}$
versus $\beta$. Our NP results from section~\ref{lattice}
are shown as filled circles. They are to be compared with the TRB-PT
results denoted by empty squares. While there is a difference between
the results, it is decreasing and thus may be primarily due to
remnant $O(a^2)$ effects, which disappear in the continuum limit.
That various determinations of $Z^{\rgi}_{\tilde{m}}$ have different
numerical values can be seen from the results of \cite{dellamorte05a}
(open diamonds and triangles). In these results two different
definitions of the axial renormalisation constant have been used,
\cite{dellamorte05c}. $Z_A$ is computed when dropping certain
disconnected diagrams, while $Z_A^{\con}$ includes them. (The difference
between the two definitions is an $O(a^2)$ effect.) Using $Z^{\con}_A$
in $Z^{\rgi}_{\tilde{m}}$ leads, perhaps coincidently, to very similar
results to our NP results.

\begin{figure}[t]
   \epsfxsize=7.25cm 
      \epsfbox{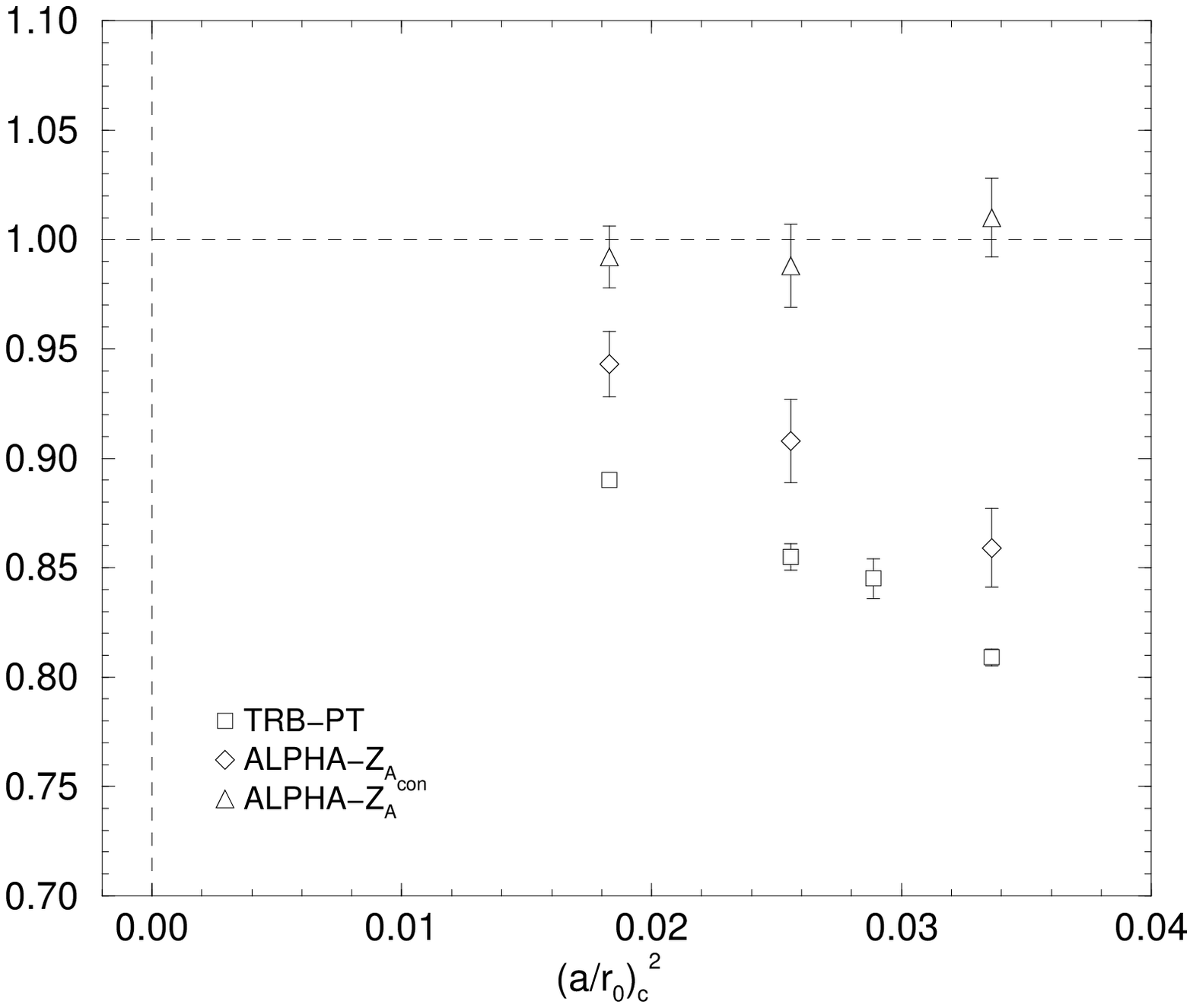}
   \vspace*{-0.25in}
   \caption{\footnotesize{
            $R_{\tilde{m}}^{X}$ versus $(a/r_0)_c^2$ for $X = \mbox{TRB-PT}$ 
            (open squares), $X = \mbox{ALPHA-$Z_A$}$ (open diamonds)
            and $X = \mbox{ALPHA-$Z_A^{\con}$}$ (open triangles).}}
   \vspace*{-0.20in}
   \label{fig_Zm_rgi_ratio_051026}
\end{figure}

Investigating the possibility of $O(a^2)$ differences a little
further, we note that if we have two definitions of $Z^{\rgi}_{\tilde{m}}$
then if both are equally valid, forming the ratio should yield
\begin{equation}
   R_{\tilde{m}}^X 
     \equiv { Z^{\rgi(X)}_{\tilde{m}} \over  Z^{\rgi}_{\tilde{m}} }
     = 1 + O(a^2) \,,
\end{equation}
where $Z^{\rgi}_{\tilde{m}}$ is the result of section~\ref{lattice}
and $X$ is some alternative definition
(i.e.\ TRB-PT, ALPHA-$Z_A$, ALPHA-$Z_A^{\con}$ ).
In Fig.~\ref{fig_Zm_rgi_ratio_051026} we plot this ratio for these
alternative definitions.
The $r_0/a$ values used for the $x$-axis are found by extrapolating
the $r_0/a$ results to the chiral limit. This extrapolation and results
(for $(r_0/a)_c$) are given in \cite{gockeler05a}.

We see that (roughly) all three ratios extrapolate to $1$ which implies
that any of the four determinations of $Z^{\rgi}_{\tilde{m}}$ may be
used. This includes the TRB-PT result. Of course other TI determinations
might not have this property, and also their validity always has to be
checked against a NP determination, so this result here is of limited use.
It is also to be noted that different determinations can have rather
different $O(a^2)$ corrections, so a continuum extrapolation is 
always necessary.

%----------------------------------------------------------------------------

\section{RESULTS}
\label{results}

\begin{figure*}[t]
   \hspace{1.00in}
      \epsfxsize=10.00cm
         \epsfbox{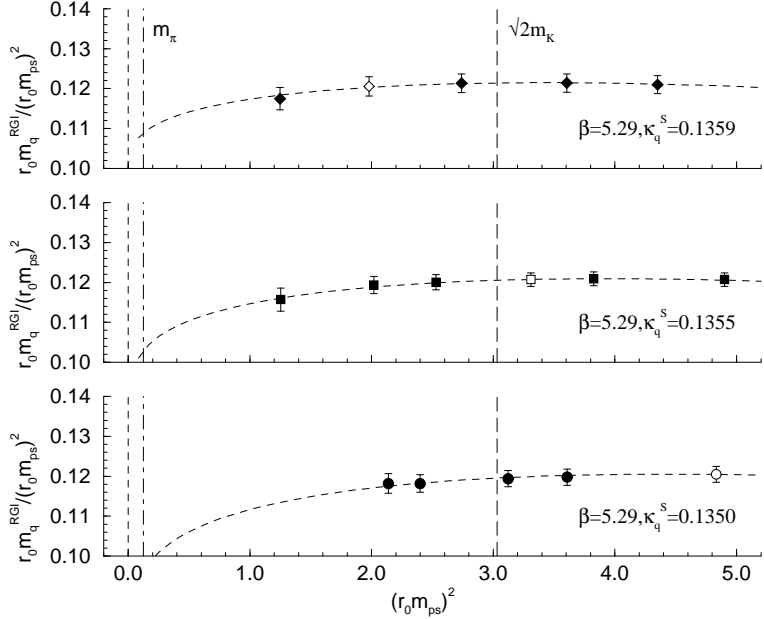}
   \vspace*{-0.35in}
   \caption{\footnotesize{
            $r_0m_q^{\rgi}/(r_0m_{ps})^2$ against $(r_0m_{ps})^2$,
            together with a fit using eq.~(\protect\ref{strange_degenerate})
            for $\beta = 5.29$. Filled points represent valence quark results
            while unfilled points are the sea quark results.
            The dashed line (labelled `$\sqrt{2}m_K$')
            represents a fictitious particle composed of two
            strange quarks, which at LO $\chi$PT is given
            from eq.~(\protect\ref{strange_general}) by
            $\sqrt{(r_0m_{K^+})^2 + (r_0m_{K^0})^2 - (r_0m_{\pi^+})^2}$,
            while the dashed-dotted line (labelled `$m_\pi$') representing
            a pion with mass degenerate $u/d$ quark is given
            by $r_0m_{\pi^+}$.}}
%   \vspace*{-0.25in}
   \label{fig_r0mps2_r0mqRGIor0mps2_b5p29_3pic_051103}
\end{figure*}

Armed with $Z_m^{\rgi}$, we can now find $r_0m_q^{\rgi}$ and hence the
ratio $r_0m_q^{\rgi}/(r_0m_{ps})^2$, using the values
of $r_0/a$ given in \cite{gockeler05b}. In
Fig.~\ref{fig_r0mps2_r0mqRGIor0mps2_b5p29_3pic_051103} we plot this ratio
(against $(r_0m_{ps})^2$) for $\beta = 5.29$. Using
eq.~(\ref{strange_general}) to eliminate $c_a^{\rgi}$ in
eq.~(\ref{strange_degenerate}) in favour of $c^{\rgi}_{a^\prime}$ where
\begin{equation}
   c^{\rgi}_{a^\prime} = 
      { r_0m_s^{\rgi} \over
        (r_0m_{K^+})^2 + (r_0m_{K^0})^2 - (r_0m_{\pi^+})^2 } \,,
\end{equation}
gives $r_0m_s^{\rgi}$ directly%
\footnote{This is preferable to first determining
$c_a^{\rgi}$ and $c_i^{\rgi}$, $i=b,c,d$ by using
eq.~(\ref{strange_degenerate}) and then substituting in 
eq.~(\ref{strange_general}) as the direct fit reduces the final 
error bar on $r_0m_s^{\rgi}$.}
to NLO in our fit function.

We have restricted the quark masses to lie in the range $(r_0m_{ps})^2 < 5$,
which translates to $m_{ps} \lsim 880\, \mbox{MeV}$, which is hopefully
within the range of validity of low order $\chi$PT results. (Using
$r_0/a$, rather than their chirally extrapolated values for example,
tends to give less variation in the ratio $r_0m_q^{\rgi}/(r_0m_{ps})^2$
so we expect LO $\chi$PT to be markedly dominant.) Varying this range
from $4$ to $6$ and higher gave some idea of possible systematic errors.
Thus finally, for each $\beta$-value we have determined $r_0m_s^{\rgi}$
and can now perform the last extrapolation to the continuum limit.

Our derivation so far, although needing a secondary quantity such as
$r_0/a$ for a unit, depends only on lattice quantities. Only at the
last stage, with our direct fit did we need to give a physical scale
to this unit. A popular choice is $r_0 = 0.5\, \mbox{fm}$.
However there are some uncertainties in this value; our derivation
using the nucleon gave $r_0 = 0.467\, \mbox{fm}$ and so to give some
idea of scale uncertainties, we shall also consider this value.
(The main change when changing the scale comes from the $r_0$s in
eq.~(\ref{strange_general}), as $m^{\rgi}_s \propto r_0$, while
changes in $\Delta Z^{\msbar}_m$ are only logarithmic.)

\begin{figure}[t]
   \begin{center}
      \epsfxsize=7.00cm
         \epsfbox{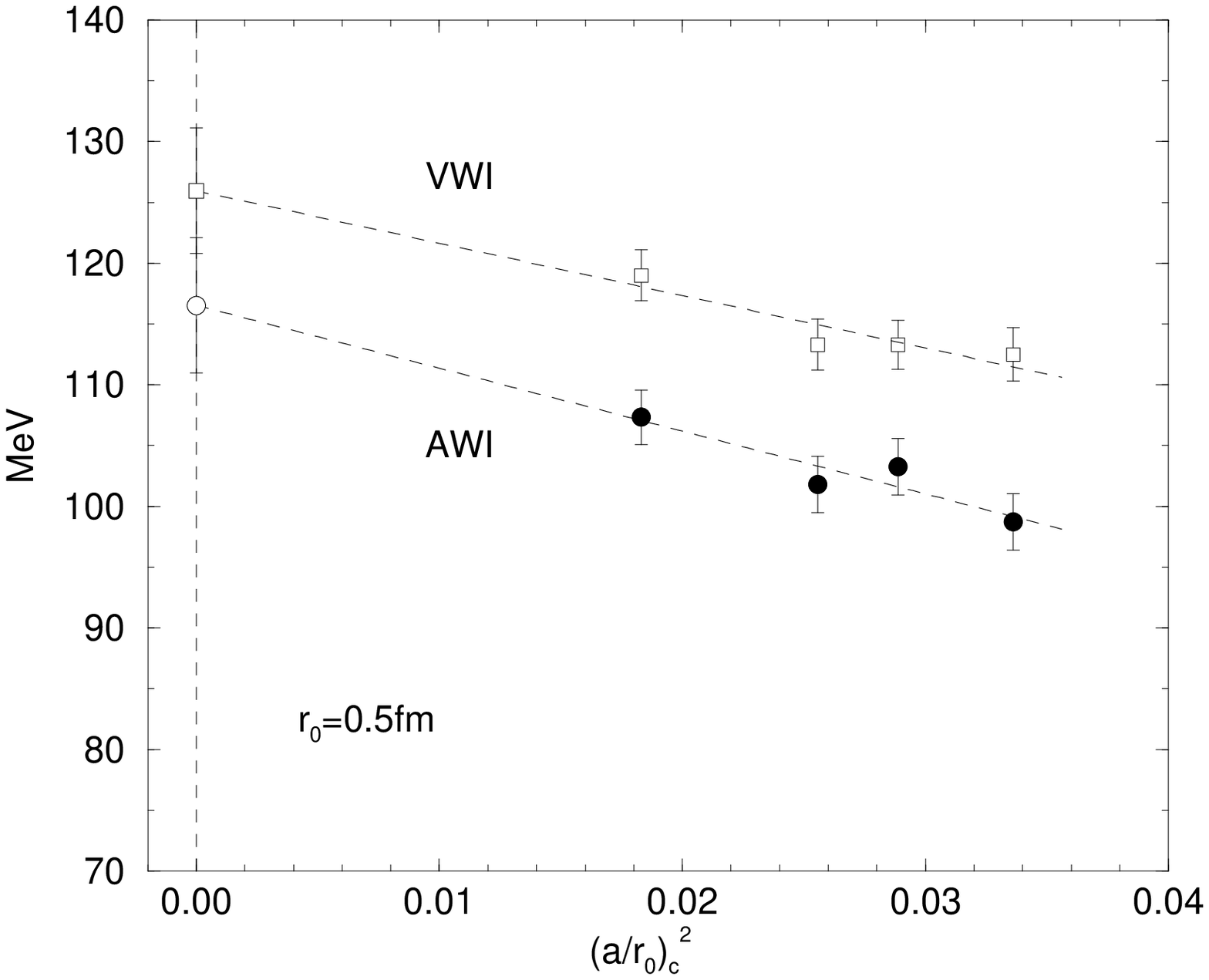}
   \end{center}
   \vspace*{-0.35in}
   \caption{\footnotesize{
            Results for $m_s^{\msbar}(2\,\mbox{GeV})$ (filled circles)
            versus the chirally extrapolated values of $(a/r_0)^2$ (as
            given in \protect\cite{gockeler05b}) together with a linear
            extrapolations to the continuum limit. For comparison, 
            we also give our previous result using the VWI,
            \protect\cite{gockeler04a} (open squares).}}
   \vspace*{-0.25in}
   \label{fig_msMSbar_AWI+VWI_LO+NLO_051103}
\end{figure}

Using the value for $[\Delta Z_m^{\msbar}( 2 \, \mbox{GeV})]^{-1}$ 
obtained in  section~\ref{rgi} from Fig.~\ref{fig_Del_Zm_MSbar_nf2_muolam}
to convert $m_s^{\rgi}$ to $m_s^{\msbar}(2\,\mbox{GeV})$ gives
the results shown in Fig.~\ref{fig_msMSbar_AWI+VWI_LO+NLO_051103}.
Also shown is an 
extrapolation to continuum limit. We finally obtain the result
\begin{eqnarray}
   \lefteqn{m_s^{\msbar}(2\,\mbox{GeV}) =}
           & &                        \label{ms_MeV_results}        \\
           & & \left\{ \begin{array}{lll}
                          117(6)(4)(6)\,\mbox{MeV} & \mbox{for} &
                          r_0 = 0.5 \,\mbox{fm}   \\
                          111(6)(4)(6)\,\mbox{MeV} & \mbox{for} &
                          r_0 = 0.467 \, \mbox{fm}\\
                       \end{array}
               \right. \,,
                                                     \nonumber
\end{eqnarray}
where the first error is statistical. The second error is systematic
$\sim 3\,\mbox{MeV}$ estimated by varying the fit interval for
$(r_0m_{ps})^2$. We take a further systematic error on these results
as being covered by the different $r_0$ values of about
$\sim 6\, \mbox{MeV}$. This is to be compared to
our previous result using the VWI, \cite{gockeler04a},
which gave results of $126(5)\, \mbox{MeV}$, $119(5)\, \mbox{MeV}$
for $r_0 = 0.5\, \mbox{fm}$ and $0.467\, \mbox{fm}$ respectively.
These results and extrapolation are also shown in
Fig.~\ref{fig_msMSbar_AWI+VWI_LO+NLO_051103}.

%----------------------------------------------------------------------------

\section{COMPARISONS}
\label{comparisons}

\begin{figure*}[t]
   \hspace{1.00in}
      \epsfxsize=10.00cm \epsfbox{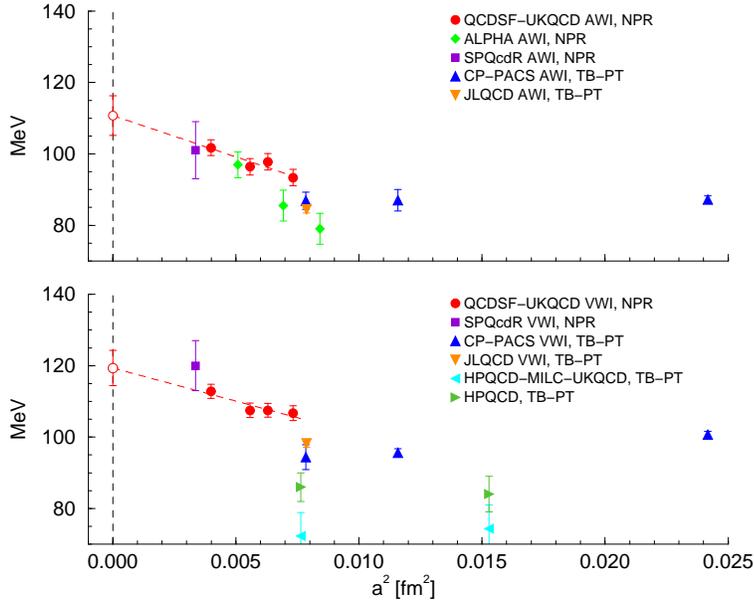}
   \vspace*{-0.35in}
   \caption{\footnotesize{
            Results for $m_s^{\msbar}(2\,\mbox{GeV})$ versus
            $a^2\,\mbox{fm}^2$ using the AWI (upper plot) and VWI
            (lower plot) methods. The results are presented with the
            collaborations preferred units and scales. Circles
            (together with a linear continuum extrapolation)
            are from this work and \cite{gockeler04a}; diamonds from
            \cite{dellamorte05b}; squares from \cite{becirevic05a};
            up triangles from \cite{alikhan01a}; down triangles from
            \cite{aoki02a}; left triangles from
            \cite{aubin04a}; right triangles from \cite{mason05a}.
            NPR denotes non-perturbative renormalisation, while TB-PT
            denotes tadpole-improved boosted perturbation theory.
            \cite{aubin04a,mason05a} are for $n_f = 2+1$ flavours;
            the other results are all for $n_f=2$ flavours.}}
   \label{fig_comparison_msbar_051116}
\end{figure*}

It is also useful to compare our results with the results from other groups.
In Fig.~\ref{fig_comparison_msbar_051116} we show some results for
$n_f=2$ and $n_f=2+1$ flavours (keeping the aspect ratio approximately
the same as in Fig.~\ref{fig_msMSbar_AWI+VWI_LO+NLO_051103}.
A variety of actions, renormalisations, units and scales have been used
(so the results have been plotted in physical units using the authors
prefered values). In particular the HPQCD-MILC-UKQCD \cite{mason05a}
and HPQCD \cite{aubin04a} collaborations use improved staggered fermions.
These fermions having a (remnant) chiral symmetry are in the same situation
as overlap/domain wall fermions where there is no distinction between
VWI and AWI quark masses; the bare quark mass in the Lagrangian
simply needs to be renormalised.

As seen earlier in section~\ref{comparison_ti} it is noticeable
that the (tadpole improved) perturbative results lie lower
than the non-perturbatively renormalised results.
Also results with $a \lsim 0.09\,\mbox{fm}$
(i.e.\ $a^2 \lsim 0.008\,\mbox{fm}^2$) appear to be reasonably consistent
with each other (this is more pronounced for the AWI results than for
the VWI results). While results for $a \lsim 0.09$ show some lattice
discretisation effects, using results at larger lattice spacings
seems to give a fairly constant extrapolation to the continuum limit.
A similar effect has also been seen elsewhere, for example in the 
determination of $r_0\Lambda^{\msbar}$ for $n_f=0$ flavours,
\cite{gockeler05b}, where coarse lattices also show this characteristic
flattening of the data.

Finally, we compare these numbers with results from the QCD sum rule
approach. A recent review of results from this method is given in
\cite{narison05a}, citing as a final result
$m_s^{\msbar}(2\,\mbox{GeV}) = 99(28)\,\mbox{MeV}$.
This covers the lattice results in Fig.~\ref{fig_comparison_msbar_051116}.

%----------------------------------------------------------------------------

\section{CONCLUSIONS}
\label{conclusions}

In this article we have estimated the strange quark mass for $2$-flavour
QCD and found the result in eq.~(\ref{ms_MeV_results}), using
$O(a)$-improved clover fermions and taking into consideration
non-perturbative (NP) renormalisation, the continuum extrapolation
of the lattice results and the use of chiral perturbation theory.
The NLO chiral perturbation theory yields a correction of about $5\%$
to the LO result, and the relevant low energy constants
are in rough agreement with the phenomenological values.

In conclusion, although there is a spread of results, it would
seem that the unquenched strange quark mass determined here is
not lighter than the quenched strange quark mass and lies in the range of
$100$ -- $130 \, \mbox{MeV}$.

%----------------------------------------------------------------------------

\section*{ACKNOWLEDGEMENTS}

The numerical calculations have been performed on the Hitachi SR8000 at
LRZ (Munich), on the Cray T3E at EPCC (Edinburgh)
%under PPARC grant PPA/G/S/1998/00777,
\cite{allton01a}, on the Cray T3E at NIC (J\"ulich) and ZIB (Berlin),
as well as on the APE1000 and Quadrics at DESY (Zeuthen).
We thank all institutions.
This work has been supported in part by
the EU Integrated Infrastructure Initiative Hadron Physics (I3HP) under
contract RII3-CT-2004-506078
and by the DFG under contract FOR 465 (Forschergruppe
Gitter-Hadronen-Ph\"anomenologie).

%----------------------------------------------------------------------------

%----------------------------------------------------------------------------

\end{document}